\documentclass[pra,aps,twocolumn]{revtex4}
\usepackage{hyperref}
\usepackage{amsfonts}
\usepackage{amssymb}
\usepackage[dvips]{graphicx}

\usepackage[T1]{fontenc}
\usepackage[latin2]{inputenc}

\usepackage{amsmath}
\usepackage{bbm}

\def\duzomniejsze{<\kern-.7mm<}
\def\duzowieksze{>\kern-.7mm>}

\def\textbf#1{{\bf #1}}
\def\beq{\begin{equation}}
\def\eeq{\end{equation}}
\def\be{\begin{equation}}
\def\ee{\end{equation}}
\def\ben{\begin{eqnarray}}
\def\een{\end{eqnarray}}
\def\beqa{\begin{eqnarray}}
\def\eeqa{\end{eqnarray}}
\def\eea{\end{array}}
\def\bea{\begin{array}}
\newcommand{\bei}{\begin{itemize}}
\newcommand{\eei}{\end{itemize}}
\newcommand{\bee}{\begin{enumerate}}
\newcommand{\eee}{\end{enumerate}}

\def\hcal{{\cal H}}

\def\tr{{\rm Tr}}

\def\>{\rangle}
\def\<{\langle}
\def\blacksquare{\vrule height 4pt width 3pt depth2pt}

\def\ot{\otimes}

\def\dt#1{{{\kern -.0mm\rm d}}#1\,}
\def\pnorm{p}

\def\sigalpe{{\sigma_\alpha'}^{\kern-.7mm E}}
\def\sigalpb{{\sigma_\alpha'}^{\kern-.7mm B}}

\newtheorem{lemma}{Lemma}

\newtheorem{theorem}{Theorem}
\newtheorem{proposition}{Proposition}
\newtheorem{definition}{Definition}
\newtheorem{fact}{Fact}

\def\bep{\begin{proposition}}
\def\eep{\end{proposition}}
\def\bel{\begin{lemma}}
\def\eel{\end{lemma}}

\def\bet{\begin{theorem}}
\def\eet{\end{theorem}}
\def\bed{\begin{definition}}
\def\eed{\end{definition}}
\def\bef{\begin{fact}}
\def\eef{\end{fact}}

\begin{document}


\title{Constructive counterexamples to additivity of minimum output 
R\'enyi entropy of quantum channels for all $p>2$}

\author{
  Andrzej Grudka${}^1$,
  Micha\l{} Horodecki${}^2$ and
  \L{}ukasz Pankowski${}^{2,3}$}

\affiliation{
  ${}^1$Faculty of Physics, Adam Mickiewicz University, 61-614
  Pozna\'{n}, Poland\\
  ${}^2$Institute of Theoretical Physics and Astrophysics,
  University of Gda\'nsk, 80-952 Gda\'nsk, Poland\\
  ${}^3$Institute of Informatics, University of Gda\'nsk, 80-952 Gda\'nsk, Poland}


\begin{abstract}
  We present a constructive example of violation of additivity of
  minimum output R\'enyi entropy for each $p>2$. The example is
  provided by antisymmetric subspace of a suitable dimension. We
  discuss possibility of extension of the result to go beyond $p>2$
  and obtain additivity for $p=0$ for a class of entanglement breaking
  channels.
\end{abstract}

\maketitle

\section{Introduction}

The question whether the minimal output entropy of quantum channels is
additive had been open for quite a long time.  While the most
interesting case is when the entropy is the von Neumann one, the more
general R\'enyi $\pnorm$-entropies (or equivalently $p$-norms) were
also studied. Several additivity results had been first obtained for
particular classes of channels, including unital qubit channels
\cite{King01-qubit} and entanglement breaking channels
\cite{Shor-break} for the von Neumann entropy
(cf. \cite{BrandaoH-Hastings2009} for a more complete list). The first
counterexample was obtained for $p\geq 4.79$
\cite{HolevoWerner-additivity2002}, for the so called Werner-Holevo
channel. Subsequently Winter \cite{Winter-additivity2007} proved
nonadditivity for all $p>2$ which was pushed by Hayden
\cite{Hayden-additivity2007} until all $p>1$ (See also
\cite{Hayden-Winter-additivity2008}). Finally Hastings has shown
nonadditivity for $p=1$ which is the von Neumann entropy case
\cite{Hastings-additivity2008} (see
\cite{FukudaKM-Hastings2009,BrandaoH-Hastings2009} 
in this context).
For a concise review of additivity problem (which does not yet include
Hastings' result) see \cite{Ruskai2007-problems}.

The counterexamples to additivity, apart from that of
\cite{HolevoWerner-additivity2002} are nonconstructive: they hold for
randomly picked channels. The purpose of this paper is to provide
explicit counterexamples for all $p>2$. We will have to increase
dimension towards infinity, as $p$ approaches $2$, similarly, as in
\cite{Winter-additivity2007}. Our channels have input dimension ${d
  \choose 2}$ and output one -- $d$. The channel with $d=3$ is
actually the Werner-Holevo channel.  Following Hayden, we use the
picture of Stinespring dilation, i.e. instead of channels, we work
with equivalent subspaces in bipartite systems. The subspace which
produces required counterexample is simply the antisymmetric
subspace. We then discuss possible extensions of the results. Finally,
as a separate result, we provide a class of entanglement breaking
channels satisfying additivity for $p=0$.

\section{Constructive counterexamples for $p>2$}

The question of additivity of minimum output entropy of a quantum
channel is the following.  Consider two channels $\Lambda_1$ and
$\Lambda_2$, and fix $p$ for quantum R\'enyi entropy $S_\pnorm={1\over
  1-\pnorm}\log \tr \rho^\pnorm$. Define minimum output entropy of a
channel $\Lambda$ as
\be
S_\pnorm^{\min}(\Lambda) =\min_\psi S_\pnorm(\Lambda(|\psi\>\<\psi|)),
\ee
where the minimum runs over all pure input states.
Then we have additivity if
\be
S_\pnorm^{\min}(\Lambda_1\ot \Lambda_2)=
S_\pnorm^{\min}(\Lambda_1)+S_\pnorm^{\min}(\Lambda_2)
\ee
while additivity is violated, when
\be
S_p^{\min}(\Lambda_1\ot \Lambda_2)<
S_p^{\min}(\Lambda_1)+S_p^{\min}(\Lambda_2)
\ee
(the inequality $"\geq"$ always holds).

As explained in \cite{Hayden-additivity2007}, using Stinespring
dilation of channel, it is easy to see that the problem of additivity
of output minimal R\'enyi entropy is equivalent to the following
question. We start with bipartite Hilbert spaces $\hcal_{A:B}$,
$\hcal_{A'B'}$, and consider subspaces $\hcal_{(1)}\subset
\hcal_{A:B}$, and $\hcal_{(2)}\subset \hcal_{A':B'}$ with
corresponding projectors $P_{AB}^{(1)}$ and $P_{A'B'}^{(2)}$.  Then
the question is whether for any state $\psi$ from the subspace
$\hcal_{(1)}\ot\hcal_{(2)}$ with corresponding projection
$P_{AB}^{(1)}\ot P_{A'B'}^{(2)}$
we have
\be
S_\pnorm(\rho_{BB'}) \geq \min_{\psi_{AB}\in\hcal_{(1)}} S_\pnorm(\rho_B) +
\min_{\psi_{A'B'}\in\hcal_{(2)}} S_\pnorm(\rho_{B'}),
\label{eq:s_min}
\ee
where $\rho_{BB'}$ etc. are reduced density matrices of the
corresponding pure states.  In other words, the question is whether
minimal entanglement of subspace measured by R\'enyi entropy of
reduction \cite{BDSW1996} is additive.

Therefore to construct a counterexample, one should find such
projectors $P_{AB}^{(1)}$ and $P_{A'B'}^{(2)}$, that

\bei
\item[1)] all states from the corresponding subspaces are highly
  entangled in cut $A:B$ and $A':B'$ respectively

\item[2)] there exists a state with support $P_{AB}^{(1)} \ot P_{A'B'}^{(2)}$
which has small entanglement in cut $AA':BB'$.
\eei

As proposed in \cite{Winter-additivity2007} it is convenient to take
$\hcal_{AB}$ isomorphic to $\hcal_{A'B'}$, and
$P_{A'B'}^{(1)}=\overline{P_{AB}^{(2)}}$, where the bar denotes
complex conjugation in standard product basis.  The trial joint state
is a maximally entangled state in $AB:A'B'$ cut, with local density
matrices proportional to the projections.

There is a lemma by Hayden \cite{Hayden-additivity2007}, which
provides bound for entanglement of such a state in $AA':BB'$ cut:
\begin{lemma}
\label{lem:Hayden}
Consider projector $P=\sum_{i=1}^k |\psi_i\>\<\psi_i|$ on Hilbert
space $\hcal_A\ot \hcal_B$, $k\leq d_A d_B$, with $d_A,d_B$ -
dimensions of $\hcal_A,\hcal_B$.  Denote
\be
\psi^+(P)={1\over \sqrt k}\sum_{i=1}^k |\psi_i\>_{AB}|\psi_i^*\>_{A'B'},
\ee
where $A'$ and $B'$ are isomorphic with $A$ and $B$, ${}^*$ is complex
conjugation in the standard product basis of $\hcal_{A'}\ot
\hcal_{B'}$. Then the square of maximal Schmidt coefficient $a_{\max}$
in cut $AA':BB'$ satisfies
\be
\lambda_{\max}\equiv a_{\max}^2\geq {\dim P \over d_A d_B}.
\ee
\end{lemma}

We shall now take $P_{AB}=P_{A'B'}=P_a$ where $P_a$ is a projector
onto the antisymmetric subspace. Since $P_a=P_a^*$ we can apply the
above lemma to get
\be
\lambda_{\max} \geq {1\over 2} {d-1\over d}.
\ee
From this we obtain that
\ben
&&S_\pnorm(\rho_{BB'})=
{1\over 1-\pnorm} \log \tr \sum_i\lambda_i^\pnorm \geq \nonumber \\
&&\leq  {\pnorm\over 1-\pnorm} \log (\lambda_{\max})\leq
{\pnorm \over 1-\pnorm} \biggl(\log{d-1\over d} -1\biggr).
\label{eq:s_alpha}
\een

To bound $S_\pnorm (\rho_B)$ we note that arbitrary vector from
antisymmetric subspace is more entangled than a two qubit maximally
entangled state, in the sense of majorization
\cite{Nielsen-pure-entanglement}.  Namely, we have
\begin{proposition}
  Any vector from antisymmetric subspace satisfies $a_{\max}^2\leq
  1/2$, where $a_{\max}$ is maximal Schmidt coefficient of the vector.
\end{proposition}
{\bf Proof}.
We shall use the following simple fact (see
e.g. \cite{PankowskiPHH-npt2007}): for any Hilbert space $\hcal$, its
subspace $\hcal'$ and arbitrary vector $\phi\in\hcal$ we have
\be
\sup_{\psi\in \hcal'}|\<\psi|\phi\>|^2
= \<\phi | P_{\hcal'}|\phi\>.
\label{eq:max-proj}
\ee
where $P_{\hcal'}$ is the projector onto the subspace $\hcal'$.  Thus
it is enough to show that $\sup_{\psi\ot\phi}
\<\psi\ot\phi|P_{a}|\psi\ot\phi\>\leq 1/2$.  Writing $P_{as}={I-V\over
2}$ where $V$ is swap operator and using $\tr (A B^\Gamma)=\tr
(A^\Gamma B)$ where $\Gamma$ denotes the partial transpose, we obtain
the required inequality.  The proposition implies that
\be
S_\pnorm(\rho_B)\geq 1,
\ee
for any $\pnorm$.

Thus we obtain violation of additivity, if
\be
S_\pnorm(\rho_{BB'})< 2 S_\pnorm(\rho_B)
\ee
i.e.
\be
{\pnorm \over \pnorm-1} \log \biggl( {d\over d-1} +1\biggr) < 2.
\ee
Since for any $\pnorm<2$ we have  ${\pnorm \over \pnorm-1} < 2$, then
for any $\pnorm<2$, if we can choose $d$ large enough, this inequality
is satisfied. \blacksquare

\section{Towards generalizations}

The simplicity of the provided counterexample is due to the fact, that
the lower bound for entropy for a single subspace (e.g. $P_{AB}$) is
given by majorization (cf. \cite{Nielsen-pure-entanglement}), a
condition, which is easy to check, while in the Hastings' proof of
non-additivity of minimal von Neumann entropy, the lower bound was the
hardest part. Here we shall discuss possible ways to strengthen the
result, by e.g. finding some other subspaces with similar majorization
property. We shall first provide a lemma, which puts some
limitations on the method.  Let us introduce the following
notation. For a given subspace $\hcal_{(1)}$ we define 
\be
\lambda^{(1)}_{\max}=\max_{\psi_{A:B}\in\hcal_{(1)}}a^2_{\max}(\psi_{A:B}),
\ee
where $a_{\max}(\psi)$ is the largest Schmidt coefficient of
the the state $\psi$. Let also
\be
\lambda^{(1:2)}_{\max}=\max_{\psi_{AA':BB'}\in\hcal_{(1)}\ot\hcal_{(2)}}
a^2_{\max}(\psi_{AA':BB'}).
\ee
If we take such subspaces, that corresponding projectors are related
via complex conjugate, we have by Hayden's lemma \ref{lem:Hayden} that
\be
\lambda^{(1:2)}_{\max}\geq {d\over d_A d_B}
\ee
where $d=\dim \hcal_{(1)}$. Now, we will prove also that
\begin{lemma}
\label{lem:amax}
With the above notation we have
\be
\lambda^{(1)}_{\max}\geq {d\over d_A d_B}.
\label{eq:lem-amax}
\ee
\end{lemma}
{\bf Proof.}
According to the formula (\ref{eq:max-proj}) we have to provide bound for
\be
\sup_{\psi\ot\phi} \<\psi\ot\phi|P|\psi\ot\phi\>.
\ee
Obviously we have
\be
\sup_{\psi\ot\phi} \<\psi\ot\phi|P|\psi\ot\phi\>
\geq \biggl\<\<\psi\ot\phi|P|\psi\ot\phi\>\biggr\>
\ee
where $\biggl\<\cdot\biggr\>$ denotes average over product states
(obtained by applying random unitaries to a standard one). Computing
the average we obtain
\be
\sup_{\psi\ot\phi} \<\psi\ot\phi|P|\psi\ot\phi\>\geq
{\dim P\over d_A d_B}.
\ee
This ends the proof. \blacksquare

Using the above lemma it is easy to see, that if we want to base on
the estimates made in the proof, no subspace that has minimal
entanglement equal to $1$ (or, even more generally $\log k$ for
integer $k$) can lead to violation for $p\leq 2$. Indeed, suppose that
$\lambda_{\max}^{(1)}=1/k$. Then $S_{\min}^{(1)}=\log
\lambda^{(1)}_{\max}$ and by \eqref{eq:lem-amax} we have
\be
2 \log S_{\min}^{(1)}\leq 2 \log {d\over d_A d_B}
\ee
Now, if we apply the estimate  used in \eqref{eq:s_alpha}
\be
{1\over 1-\pnorm}\sum_i \lambda_i^\pnorm \leq {\pnorm\over 1-\pnorm} \lambda_{\max}
\label{eq:est-alpha}
\ee
we obtain
\be
{\pnorm\over 1-\pnorm}\log \lambda^{(1:2)}_{\max} \leq 2 \log {d \over d_A d_B}.
\ee
If we now use lemma \ref{lem:Hayden} to estimate
$\lambda^{(1:2)}_{\max}$, the inequality \eqref{eq:s_min} becomes
$\pnorm/(1-\pnorm)\geq 2$ which can be only violated for $\pnorm>2$.

There is some gap here due to two non-tight estimates: the one of
lemma \ref{lem:Hayden} and that of \eqref{eq:est-alpha}. In the case
of antisymmetric subspace, this does not enlarge the region of
violation, and it seems that we have additivity for $\pnorm\leq 2$ in
case of antisymmetric subspace. In general, however, it turns out that
if we keep the estimate of lemma \ref{lem:Hayden}, there is still
chance for violation of additivity even for the von Neumann entropy.
E.g. we get the following sufficient condition for violation of
additivity for the von Neumann entropy.

{\it Additivity is violated, if there exists $D$ and $d$ such that in
the space $C^{D}\ot C^{D}$ there exists a subspace $\hcal'$ of
dimension $d$, such that
\be
2(1-{d\over D^2}) \log D + h({d\over D^2})<2
\ee
and any vector in $\hcal'$ has entropy of entanglement
no less than $1$.}

In order to violate additivity it would be then enough to get such
subspace for which $d/D^2\to 1$ with entanglement bounded from below
by a constant.
We have checked numerically the subspaces such as the $(d-1)^2$
dimensional space constructed by Parthasarathy
\cite{Parthasarathy04-compl-ent} which does not contain any product
vector, as well as those provided in \cite{CubittMW07-rank} - largest
subspaces, which contain only vectors with Schmidt rank no less than
given value $k$. Unfortunately, the subspaces do not exhibit required
properties.

Finally, one may ask the question of multiple additivity, when one
takes more than two subspaces (e.g. the Werner-Holevo channel was
shown to be multiple-additive for $1\leq \pnorm\leq 2$ in
\cite{AlickiF-additivity2004}). We have checked numerically multiple
additivity of the antisymmetric subspaces for $\pnorm=\infty$ and
obtained evidence that the optimal entangled input is bipartite,
\mbox{i.e.} even for more copies it does not pay to use multipartite
entangled states. For even number of copies, the optimal suggested
state is just a product of bipartite maximally entangled states, while
for odd number of copies one takes product of such pairs times a
single-partite state of entanglement equal to 1. As an example of
inefficiency of multipartite states, consider $n=d(d-1)/2$ copies of
the antisymmetric subspace on $d\ot d$ system. One could think that a
totally antisymmetric state is a good candidate. However we checked
cases $d=3$ and $d=4$, and obtained that reduced density matrices are
proportional to projections, and any $p$-entropy is equal to $4$ and
$10$ respectively, which is larger than the sum of single-copy minimal
entropy, being $3$ and $6$ respectively.





\section{On R\'enyi entropy for $\pnorm=0$ for
some entanglement breaking channels}

In \cite{Shor-break} it was shown that minimum output entropy is
additive for $\pnorm=1$ if one of the channels is an entanglement
breaking channel.  This was extended to all $p\geq 1$ in
\cite{King2002-ent-break}.  Further, King has announced a proof 
for $0<p<1$ \cite{King2007-ent-break}. 
The question whether it is true for
$\pnorm=0$ is still open. Here we consider $p=0$ and a special class
of entanglement breaking channels, namely those for which the
Choi-Jamio{\l}kowski state is is proportional to
projection on a subspace spanned by so-called unextendible product
basis \cite{BennettUPBI1999} (a set of product orthogonal vectors that
cannot be extended to a larger such set). Recall, that the 
Choi-Jamio\l{}kowski state of a channel is given by
\be
(I\ot \Lambda)|\Phi_+\>\<\Phi_+|
\ee
where $|\Phi_+\>={1\over \sqrt{d}}\sum_i|i\>|i\>$.  To prove additivity, one
can e.g. show that all output states of two copies of a channel have
maximal rank.  We shall apply the following observation of
\cite{CubittHLMW-additivity2007} where first counterexample for $p=0$
was provided: If the orthogonal complement of the Choi-Jamio{\l}kowski
state for a channel $\Lambda$ has no product vector then output of the
channel $\Lambda(|\phi \rangle \langle \phi |)$ has maximal rank $d$
for every input state $| \phi \rangle$.

We shall now take the tensor product of two such Choi-Jamio{\l}kowski
states and show that its orthogonal complement does not have a product
vector.

To this end, it is enough to show, that product of two unextendible
product bases is again an unextendible product basis. However, we are
not able to show it for a general unextendible product basis.

According to Lemma 1 of \cite{BennettUPBI1999} we have to show that
there is no partition of basis states into two disjoint sets such that
the local rank $r_{i}$ of the $i$th subset is less than the dimension
$d_{i}$ of the $i$th party's Hilbert space.  We consider bipartite
case, and assume that the dimensions of Alice and Bob's Hilbert spaces
are equal, i.e., $d_{i}=d$.  Let us take $2(d-1)+1$ product states
such that any set containing $d$ of them has the following property:
it has full rank for both Alice and Bob Hilbert spaces (it is clear
that this is minimal set with this property). These states form
unextendible product base (UPB), because if we divide them into two
disjoint sets then one set has to contain at least $d$ states.

We want to prove that this construction is closed under the tensor
product. We have the following basis states $| \psi_{i} \rangle
\otimes | \phi_{i} \rangle$ for $AB$ and $| \psi_{i}' \rangle \otimes
| \phi_{i}' \rangle$ for $A'B'$. Taking the tensor product we obtain
$| \psi_{i} \rangle | \psi_{j}' \rangle \otimes | \phi_{i} \rangle |
\phi_{j}' \rangle$ and for convenience we denote each state by $| i j
\rangle$. We can divide these states into $2(d-1)+1$ partitions. The
$i$th partition contains states $|ij\rangle$ for each $j$, i.e. it has
$2(d-1)+1$ elements. Let us take the states of the first partition and
divide them into two sets. It is clear that one of these sets contains
at least $d$ states. Next we take the second partition, the third and
so on. We see that at the end one set has to contain $d$ partitions
which contain at least $d$ elements each. This results from the fact
that there are $2(d-1)+1$ partitions. Hence this set has full rank
both for Alice and Bob's Hilbert space. This completes the proof.

\section{Acknowledgement}

We thank Remigiusz Augusiak, Maciej Demianowicz, Pawe\l{} Horodecki
and Andrzej Nowakowski for numerous discussions.  The work is supported
by EC IP SCALA. Part of this work was done in National Quantum
Information Centre of Gda\'nsk.

\vfill

\bibliographystyle{apsrev4-1long}
\bibliography{rmp14-hugekey}
\end{document}